\documentclass[aps,pre,showpacs,twocolumn]{revtex4}
\usepackage{amssymb}
\usepackage{graphicx}
\usepackage{colordvi}

\begin{document}

\title{Faraday waves in quasi-one-dimensional  superfluid Fermi-Bose mixtures}
\author{F. Kh. Abdullaev,$^{1,2}$ M. \"{O}gren,$^{3}$ and M. P. S\o rensen$^{3}$ }
\affiliation{ $^{1}$Physical-Technical Institute, Uzbek Academy of
Sciences, 2-b G. Mavlyanov Street, 100084 Tashkent, Uzbekistan \\
$^{2}$Instituto de Fisica Teorica, Universidade Estadual Paulista J\'{u}lio de Mesquita Filho, R. Dr. Bento Teobaldo Ferraz 271, Barra Funda, S\~{a}o Paulo, CEP 01140-070, Brazil \\
$^{3}$Department of Applied Mathematics and Computer Science, Technical University of Denmark, 2800 Kongens Lyngby, Denmark }
\date{February 18, 2013}

\begin{abstract}
The generation of Faraday waves in superfluid Fermi-Bose mixtures in elongated
traps is investigated.
The generation of waves  is achieved by periodically changing a parameter of the system  in time.
Two types of modulations of parameters are considered: a variation of the fermion-boson scattering length and the
boson-boson scattering length. 
We predict the properties of the generated  Faraday patterns and study the parameter regions where they can be excited.
\end{abstract}

\pacs{03.75.Ss, 03.75.Kk, 05.45.-a} 
\maketitle

\section{Introduction}
Faraday waves  (FWs) are spatially periodic patterns that can be generated in a
system with a periodic variation in time of the system parameters.
Faraday waves were initially observed by Faraday for a vessel with a liquid
oscillating in the vertical direction~\cite{Faraday}.
Such a type of structure can exist in
nonlinear optical systems~\cite{Matera,Abdullaev94, Abdullaev97,ADG,Armaroli} 
where variations along the longitudinal direction of the Kerr nonlinearity can be achieved by the periodic variation of the effective cross-sectional area of the nonlinear optical fiber. 
Recently parametric resonances in modulational instability of electromagnetic waves  in photonic crystal fibers with a periodically varying diameter have been experimentally observed in Ref.~\cite{Droques}.  
Other examples are the patterns in the   Bose-Einstein condensates (BECs)  with an atomic scattering length or a radial confinement (a transverse frequency of the trap) periodically varying in time.  
For BECs the cases of one-~\cite{Staliunas,Staliunas1,Modugno2006, Kevrekidis} and two-component condensates~\cite{Indus,Nicolin}, as well
as dipolar condensates~\cite{Santos,Santos1}, have been investigated.
Related parametric amplifications in an optical lattice have been studied in Ref.~\cite{Kramer2005} and by capillary waves on the interface between two immiscible BECs in Ref.~\cite{Kobyakov2012}.
In BECs FWs were first predicted in two dimensions with a periodically varying transverse frequency of the trap~\cite{Staliunas} and later observed in an experiment with a repulsive interacting BEC, loaded into an elongated trap~\cite{Engels}.
The existence of FWs also in elongated fermionic clouds was discussed in Ref.~\cite{Ital} 
and Faraday patterns in a superfluid Fermi gas were investigated in Ref.~\cite{Tang}.

Periodic modulation of 
the coefficients of nonlinearity in the relevant mean-field equations can be achieved by variation of the atomic scattering length by the Feshbach resonance technics~\cite{Kagan,Kagan2,Inouye,Collective} or by time modulation of the transverse frequency of the trap~\cite{Staliunas}.
In the former case it is necessary to vary the external magnetic field in time near the resonant value.
The presence of a deep optical lattice has been shown to suppress the Faraday pattern generation \cite{Capuzzi}.

The purpose of this work is to investigate the mechanism of Faraday
wave generation in elongated superfluid Fermi-Bose (FB) mixtures.
Such FB mixtures have many interesting properties in comparison with the pure bosonic
case~\cite{Adhikari} and Faraday waves can be a useful tool to
measure the nonlinear properties and in particular instabilities in these systems.
Two types of atomic scattering lengths are relevant in this system:
the fermion-boson scattering length $a_{12}$ describing the scattering between the two components and the boson-boson scattering length $a_{b}$.
Variation in time of these lengths by an external magnetic field
opens the possibility of generating Faraday waves in the mixture.
The actions of these variations are different though.
In the latter bosonic
case we parametrically excite the BEC subsystem, while in the former case we
excite the bosonic and fermionic subsystems simultaneously.
Thus we can
expect different responses of temporal parametric perturbation
with various types of pattern formation. 

Strongly repulsive interacting bosons in one dimension, so-called Tonks-Girardeau gases \cite{Tonks-Girardeau},
have the same long-wavelength dynamics as noninteracting fermions \cite{1DTonks}.
Hence the results presented here can also be realized in systems with two coupled bosonic species with strong and weak intraspecies interactions.

\section{Model}
The model of a quasi-one-dimensional superfluid FB mixture is described by the following system of coupled equations for
the complex functions $\psi_{1,2}(x,t)$~\cite{Adhikari}:
\begin{eqnarray}\label{bfsys}
i\psi_{1,t} &=& -\psi_{1,xx} +  g_b |\psi_1|^2\psi_1  + g_{12} |\psi_2|^2\psi_1,\nonumber\\
i\psi_{2,t} &=& -\psi_{2,xx} +  \kappa\pi^2 |\psi_2|^4\psi_2 + g_{12} |\psi_1|^2\psi_2,
\end{eqnarray}
with components $1$~($2$) representing bosons (fermions).
In general, Bose- and Fermi subsystems are described by the Lieb-Liniger and Gaudin-Yang theories, respectively.
Here we are interested in weak Bose-Bose interactions (we consider small positive $g_b$) and attractive Fermi-Fermi interactions and the superfluid  Fermi-Bose system is described by the nonlinear Schr\"odinger-like equation~(\ref{bfsys})~\cite{Heiselberg,Bulgac,Manini,Astr,Adhikari2}.
In the BCS weak attractive coupling limit 
the  fermionic subsystem coefficient is $\kappa =1/4$, while in the molecular unitarity limit 
it is $\kappa=1/16$~\cite{Adhikari}.
Finally, for the bosonic Tonks-Girardeau limit \cite{1DTonks} 
with the components 1 (2) being a weakly (strongly) repulsive bosonic species
we have $\kappa =1$.
Furthermore, $g_{b}= 2\hbar a_b \omega_{\perp}$ is the one-dimensional coefficient of mean-field
 nonlinearity for bosons, where $a_b$ is the scattering
length and $\omega_{\perp}$ is the perpendicular frequency of the
trap. Similarly, $g_{12}$ is the interspecies interaction coefficient~\cite{Adhikari}.
The system is written in dimensionless form using the variables
\begin{displaymath}
l=\sqrt{\frac{\hbar}{m_b \omega_{\perp}}}, \ \psi = \sqrt{l} \Psi, \ t=\tau \omega_{\perp}, \ x =\frac{ X}{l}, \ g_{j} = \frac{2m_b l}{\hbar^2}G_{j},
\end{displaymath}
where $G_j$ ($j=b,12$) is the coefficient of the mean-field nonlinearity.
We also implicitly assume that $m_b = m_f$ in Eq.~(\ref{bfsys}) since such a condition can be realized approximately in the $ ^{7}$Li-$^6$Li and $^{41}$K-$^{40}$K  mixtures.
 
The mean-field nonlinearities will be varied in time as
\begin{eqnarray}\label{TDgbANDg12}
g_{12}(t) &=& g_{12}^{(0)}[1 + \alpha_{12} \cos(\Omega_{12} t)],  \nonumber\\
g_b(t) &=& g_{b}^{(0)}[1+\alpha_b \cos(\Omega_b t)].
\end{eqnarray}
Such variation can be achieved, for example, by using Feshbach
resonance techniques, namely, by variation of an external magnetic
field near a resonant value~\cite{Kagan}. 
This leads to the temporal variation
of interspecies and intraspecies scattering lengths and the respective mean-field coefficients.
The system (\ref{bfsys}) has plane-wave solutions
\begin{equation}
\psi_{1,2}=A_{1,2}\exp(i\phi_{1,2}), \ \ A_{1,2} \  \in \mathbb{R}^+,
\end{equation}
where
\begin{eqnarray}
\phi_1 &=& -g_b A_1^2 t + g_{12}A_2^2 t,\nonumber\\
\phi_2 &=& -\kappa\pi^2 A_2^4 t + g_{12}A_1^2 t.
\end{eqnarray}

\section{Modulational instability of plane waves}
Let us now study the modulational instability (MI) of nonlinear plane waves
using the linear stability analysis~\cite{ADG}.
We will look for solutions of the form
\begin{equation}
\psi_{1,2}= \left[ A_{1,2} + \delta\psi_{1,2}(x,t) \right] e^{i\phi_{1,2}(t)}, \
| \delta\psi_{1,2}|  \ll A_{1,2}. \label{ansatzLSA}
\end{equation}
Substituting these expressions into the system (\ref{bfsys}) and
linearizing, we get the  following system for $\delta\psi_{1,2}$
\begin{eqnarray}
i\delta\psi_{1,t}&-&g_b(t) A_1^2 (\delta\psi_1 +
\delta\psi_1^{\ast}) +\delta\psi_{1,xx} \nonumber\\
&+&g_{12}(t)A_1 A_2(\delta\psi_2 + \delta\psi_{2}^{\ast})=0,\nonumber\\
i\delta\psi_{2,t}&-& 2\kappa\pi^2 A_2^4 (\delta\psi_2 +
\delta\psi_2^{\ast}) +\delta\psi_{2,xx} \nonumber\\ &+&
g_{12}(t)A_1 A_2(\delta\psi_1 + \delta\psi_1^{\ast})=0.
\label{LinearSystem}
\end{eqnarray}
With $\delta\psi_1 = u+iv$ and $\delta\psi_2 = p + iq $, we now use the Fourier transforms 
$U(k,t) = \int dx u(x,t) e^{-ikx}  = \mathcal{F} \{ u(x,t) \}$ and $V(k,t) = \mathcal{F} \{ v(x,t) \}$.
Differentiation with respect to time of the remaining differential equations containing $V_t$ and $Q_t$ gives the system 
\begin{eqnarray} \label{TDODE}
V(k,t)_{tt} + \omega_1^2(t) V(k,t) &=& \varepsilon(t) Q(k,t),\nonumber\\
Q(k,t)_{tt} + \omega_2^2 Q(k,t) &=& \varepsilon(t) V(k,t),\nonumber\\
U(k,t) = k^2 \int dt \: V, \ \ P(k,t) &=& k^2 \int dt \: Q,
\end{eqnarray}
where
\begin{eqnarray}
\omega_1^2(t) = k^2 \left[ k^2 + 2g_{b}(t) A_1^2 \right],\ \ \omega_2^2 &=&  k^2(k^2 +
4\kappa \pi^2 A_2^4),\nonumber\\
\varepsilon(t) &=& 2g_{12}(t)k^2 A_1 A_2. \label{def_w1_w2}
\end{eqnarray}
The terms $g_{b}(t)$ and $g_{12}(t$) are defined in Eqs.~(\ref{TDgbANDg12}).
In the following we use the notation $\omega_1 \equiv \omega_1(t=\pi/2\Omega_b)$ and $\varepsilon_0 \equiv \varepsilon(t=\pi/2\Omega_{12})$.
Hence, by solving the coupled equations for $V$ and $Q$ for given $k$, all the components of $\delta\psi_{1,2}$ can in principle be obtained by the inverse Fourier transform.

We consider the case of a FB mixture with constant system parameters.
Looking for solutions $V$ and $Q$ with a time dependence of the form $\exp(\pm i\Omega t)$, we obtain the dispersion relation of the modulations
\begin{equation}
\Omega^2_{1,2} = \frac{\omega_1^2 + \omega_2^2}{2} \pm \sqrt{\frac{ \left( \omega_2^2 - \omega_1^2 \right)^2}{4} +\varepsilon_0^2} \label{Omega_1and2}
\end{equation}
such that in the weak-coupling limit $\varepsilon_0 \ll  \omega_2^2 -  \omega_1^2 $, 
\begin{equation}
\Omega^2_{1}  \rightarrow \omega_1^2 - \frac{ \varepsilon_0^2}{ \omega_2^2 -  \omega_1^2 }, \  \ \Omega^2_{2} \rightarrow \omega_2^2 + \frac{ \varepsilon_0^2}{ \omega_2^2 -  \omega_1^2 }, \label{Omega_1and2_limit_1}
\end{equation}
and in the limit of approaching frequencies $\omega_2 \rightarrow \omega_1$, 
\begin{equation}
\Omega^2_{1}  \rightarrow \omega_1^2 - \varepsilon_0, \  \ \Omega^2_{2} \rightarrow \omega_2^2 +  \varepsilon_0.  \label{Omega_1and2_limit_2}
\end{equation}

The stability condition (for any $k$) is obtained by requiring $\Omega_{1,2}$ in Eq.~(\ref{Omega_1and2}) to be real, i.e.,
\begin{equation}
g_{12}^{(0)} < \sqrt{2\pi^2\kappa g_b^{(0)} A_2^2}. \label{StabilityCondition}
\end{equation}
In the opposite case we have MI in the FB mixture~\cite{Adhikari}.

\section{Analysis of parametrically excited instabilities}

In the following we will consider the linear ordinary differential equation (ODE) model~(\ref{TDODE}) for different cases of periodic modulations (\ref{TDgbANDg12}). 
When only the interspecies interaction parameter $g_{12}$ is modulated, we
obtain a system of two coupled oscillators with a coupling parameter varying in time
\begin{eqnarray}
V_{tt} + \omega_1^2 V &=& \varepsilon(t) Q,\nonumber\\
Q_{tt} + \omega_2^2 Q &=& \varepsilon(t) V, \label{EqForInterMod}
\end{eqnarray}
where $\varepsilon(t) = \varepsilon_0 +  \alpha_{12} \varepsilon_0 \cos(\Omega_{12} t)$ according to Eq.~(\ref{def_w1_w2}).
When only the intraspecies parameter $g_b$ is modulated, we get a system of one Mathieu equation coupled to an oscillator equation
\begin{eqnarray}
V_{tt} + \omega_1^2(t) V &=& \varepsilon_0 Q,\nonumber\\
Q_{tt} + \omega_2^2 Q &=& \varepsilon_0 V, \label{EqForIntraMod}
\end{eqnarray}
where $\omega_1^2(t) =\omega_1^2+  \alpha_b \omega_1^2  \cos(\Omega_b  t)$.

\begin{figure}[tbp]
\begin{center} 
\includegraphics[height=3.4cm]{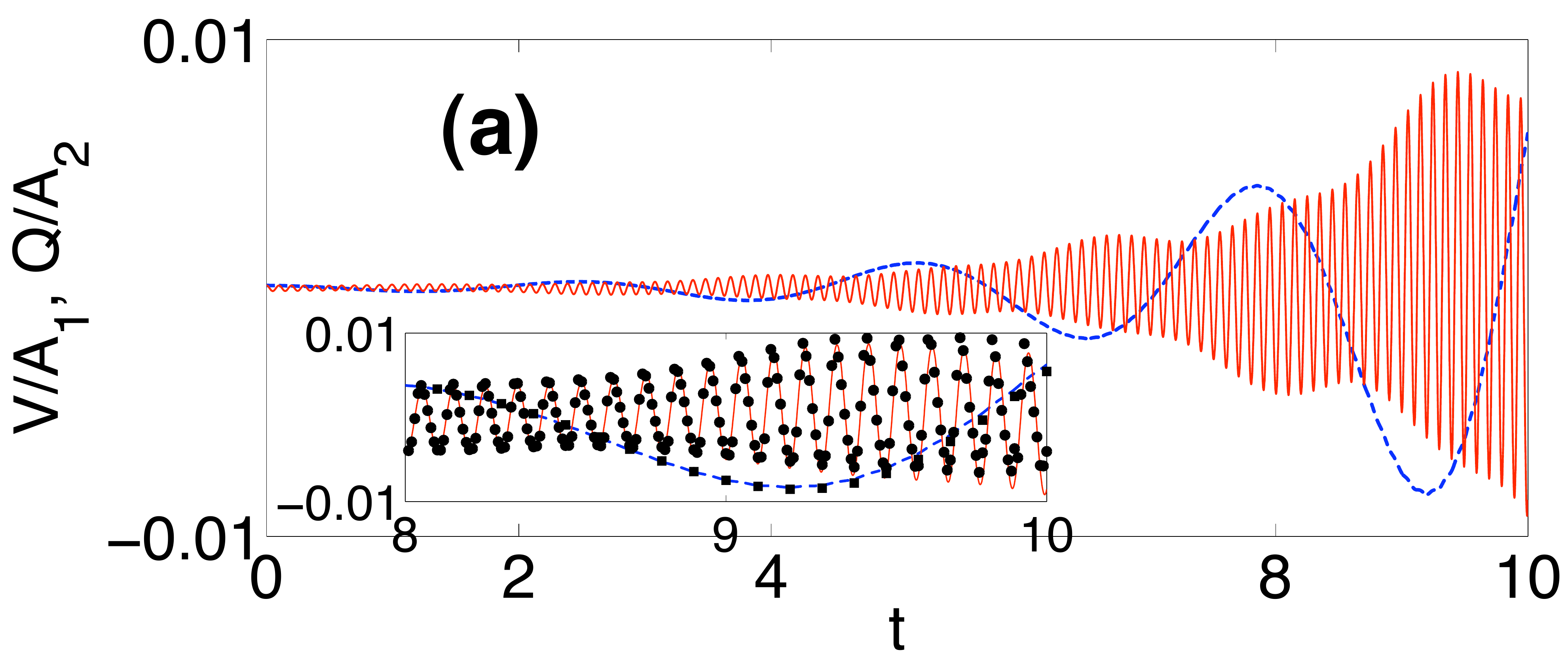}
\includegraphics[height=3.2cm]{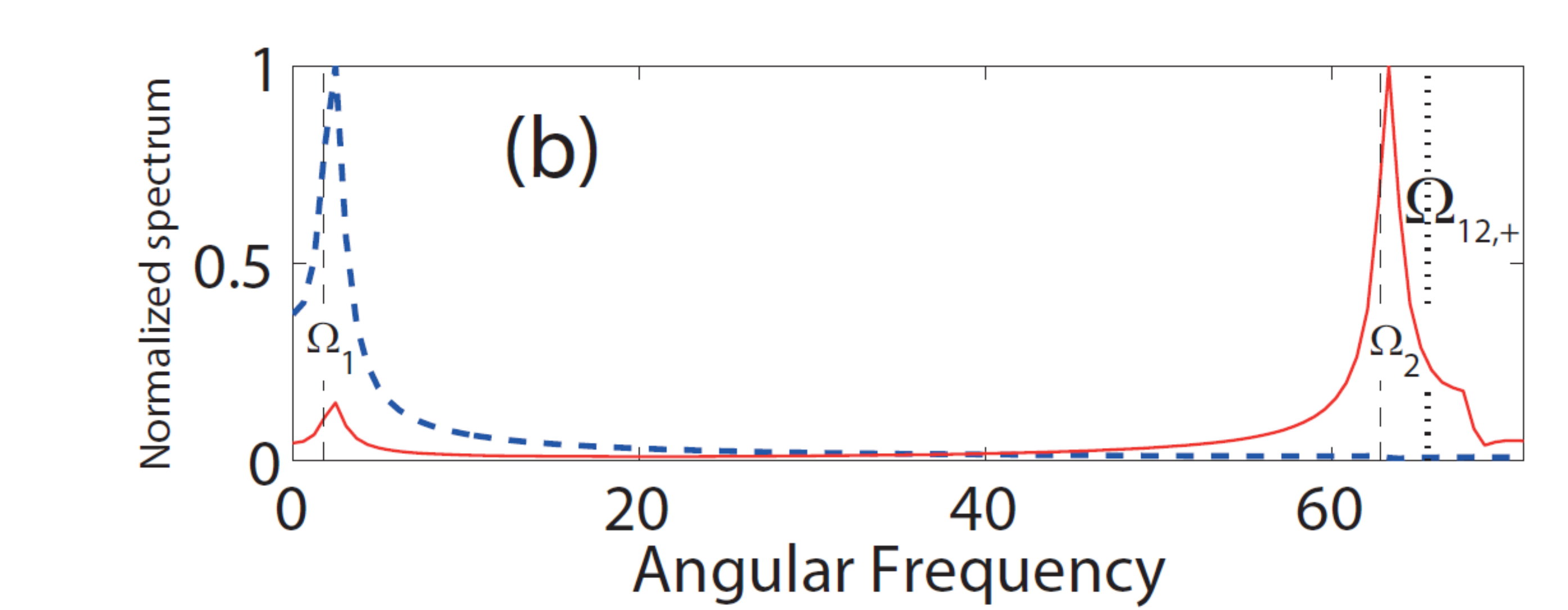}
\end{center}
\caption{(Color online) Solutions to the ODE model of Eq.~(\ref{EqForInterMod}). 
Blue dashed curves are for $V$ (bosons) and red thin curves are for $Q$ (fermions). 
(a) Amplitudes growing exponentially in time. 
The inset shows a comparison with the corresponding PDE data ($x=0$ slices of Fig.~\ref{Fig2}) for selected times.
The PDE data are shown with black squares for bosons and dots for fermions. 
(b)~Spectrum for the modulation frequency $\Omega_{12,+}= 65.5 $ for the wave number $k=1$ (see the text) and with $\Omega_{1}$ and  $\Omega_{2}$  from Eq.~(\ref{Omega_1and2}), which agrees well with Eq.~(\ref{Omega_1and2_limit_1}) in this regime. 
The initial conditions are $V/A_1=Q/A_2= 10^{-4}$ ($V_t=Q_t=0$) and the  parameters are $\alpha_{12}=0.25, \ A_1=\sqrt{300}, \ A_2=\sqrt{20}, \ g_b^{(0)} = 0.01, \ g_{12}^{(0)} = 0.8$, and $ \kappa =1/4$.}
\label{Fig1}
\end{figure}

\begin{figure}[tbp]
\begin{center} 
\includegraphics[height=2.95cm]{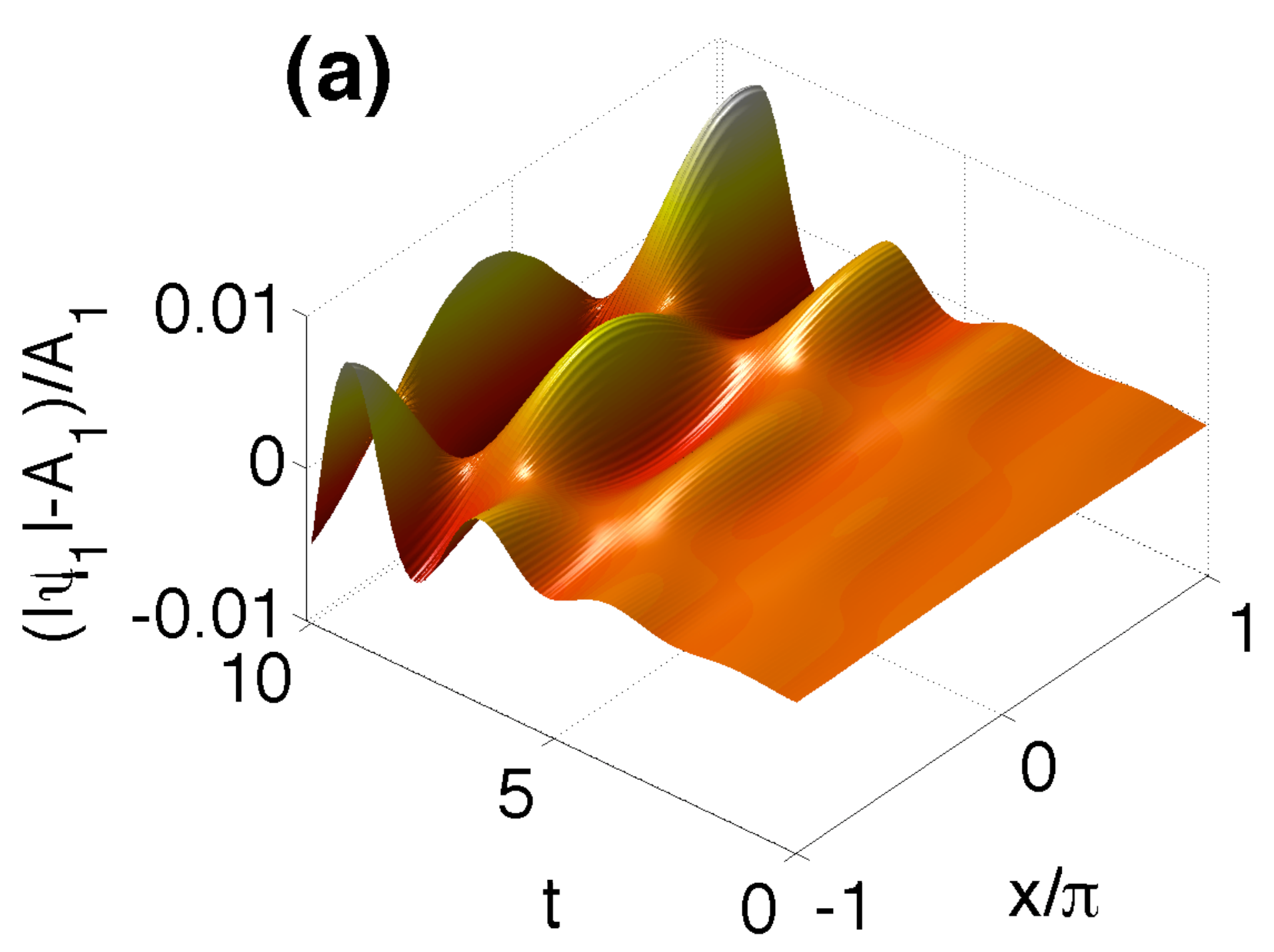}~~
\includegraphics[height=2.95cm]{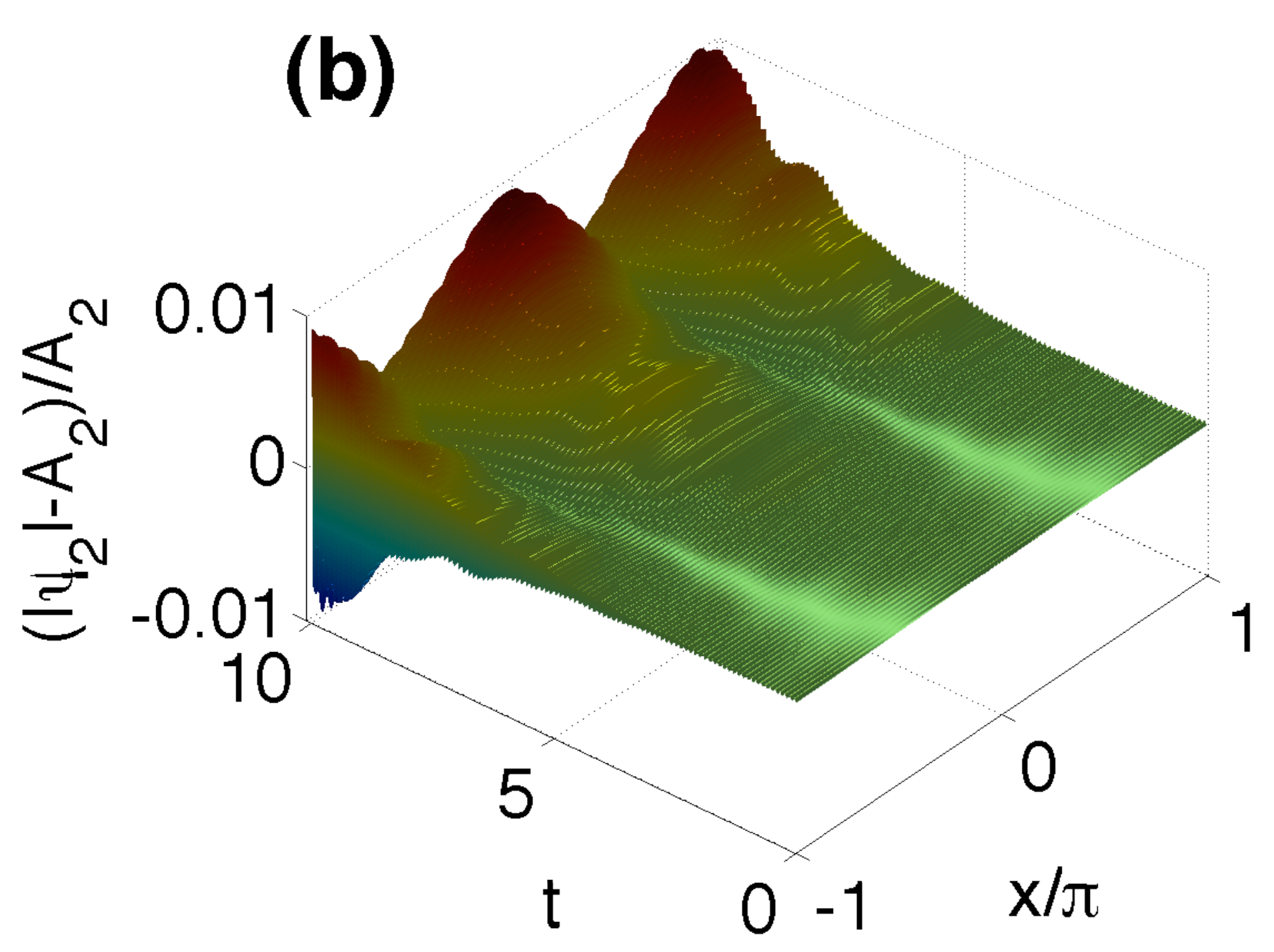}
\end{center}
\caption{(Color online) Solutions to the full PDE model of Eq.~(\ref{bfsys})  for driven coupling between fermions and bosons with the modulation frequency $\Omega_{12,+}= 65.5 $ ($k=1$). We plot the oscillating parts $\left( |\psi_j| -  A_j \right)/ A_j  $ for (a) $j=1$ (bosons) and (b) $j=2$ (fermions). The initial conditions are $\psi_j=A_j [1 + 10^{-4} \exp(ikx)]$ and the boundary conditions in $x$ are periodic. The parameters are the same as in Fig.~\ref{Fig1}.}
\label{Fig2}
\end{figure}

\subsection{Driven coupling between fermions and bosons}

We consider the case of a stable FB mixture according to Eq.~(\ref{StabilityCondition}) such that the MI is absent~\cite{Adhikari}.
Hence any instability is only due to parametric resonance.
To analyze resonances of the system~(\ref{EqForInterMod}) we use the multiscale  method~\cite{Nayfeh,Makhmoud}. 
The derivations are collected in the Appendix, where we conclude that  parametric resonances may occur at the two frequencies of excitations
\begin{equation}
\Omega_{12,-} = \omega_2 -\omega_1, \ \ \Omega_{12,+} =\omega_2 + \omega_1.  \label{Omega_pm}
\end{equation}
As the detailed analysis shows, the excitations for $\Omega_{12,-}$ are stable and do not lead to FWs. 
Inserting the expressions from Eq.~(\ref{def_w1_w2}) into  the second case of Eq.~(\ref{Omega_pm}), with the use of a simplified notation $a=2g_b^{(0)} A_1^2$ and $ b = 4\pi^2\kappa A_2^4$, 
we get the resonance frequencies in terms of the wave numbers of the corresponding Faraday waves
\begin{equation}
\Omega_{12,+}= k(\sqrt{k^2 + a} + \sqrt{k^2+b}). \label{Omega_k_2}
\end{equation}
We  invert Eq.~(\ref{Omega_k_2}) such that the wave numbers are 
\begin{equation}
k^2 =   \frac{ a+b - 2 \sqrt{ ab + \Omega_{12,+}^2 } }{(b-a)^2-4\Omega_{12,+}^2}\: \Omega_{12,+}^2. \label{k2_1}
\end{equation}
The spatial wavelength $L = 2 \pi /k$ of the FW obtained from Eq.~(\ref{k2_1}) is
\begin{equation}
L \left( \Omega, \kappa\right)  = \frac{2\pi }{\Omega_{12,+}}  \sqrt{ \frac{(b-a)^2-4\Omega_{12,+}^2}{ a+b - 2 \sqrt{ ab + \Omega_{12,+}^2 } }\: } . \label{L}
\end{equation}

According to the Appendix, the region of instability for $\Omega_{12,+}$ from Eq.~(\ref{Omega_pm}) is restricted by the lines
\begin{equation}
\bar{\omega}_1^2 \simeq \omega_1^2   \pm \frac{\varepsilon_0\alpha_{12}}{2}\sqrt{\frac{\omega_1}{\omega_2}}, \  \bar{\omega}_2^2 \simeq \omega_2^2 \pm \frac{\varepsilon_0\alpha_{12}}{2}\sqrt{\frac{\omega_2}{\omega_1}}, \label{widths_a}
\end{equation}
while the exponential growth rate of the amplitudes is restricted by the maximal gain
\begin{equation}
p_m \simeq \frac{\varepsilon_0 \alpha_{12}}{4\sqrt{\omega_1\omega_2}}. \label{p_A}
\end{equation}

A mathematically particularly simple case of the driven interspecies modulation is when $ \omega_1 \simeq \omega_2$. 
Introducing the symmetric $\xi_{+}(t)= (V+Q)/2)$ and antisymmetric $\xi_{-}(t)=(V-Q)/2$ combinations in the ODE~(\ref{EqForInterMod}), we can find that a set of parametric resonances at $\Omega_{12} \simeq  2\Omega_1,  \ 2\Omega_2$ exist.  
This follows directly from the fact that we then have two uncoupled Mathieu equations \cite{Abramowitz_and_Stegun} for the variables $\xi_{+}$ and $\xi_{-}$.
Note that resonances at low ($2\Omega_1$) and high  ($2\Omega_2$) frequencies occur [see Eq.~(\ref{Omega_1and2_limit_2})].

\begin{figure}[tbp]
\begin{center} 
\includegraphics[height=3.1cm]{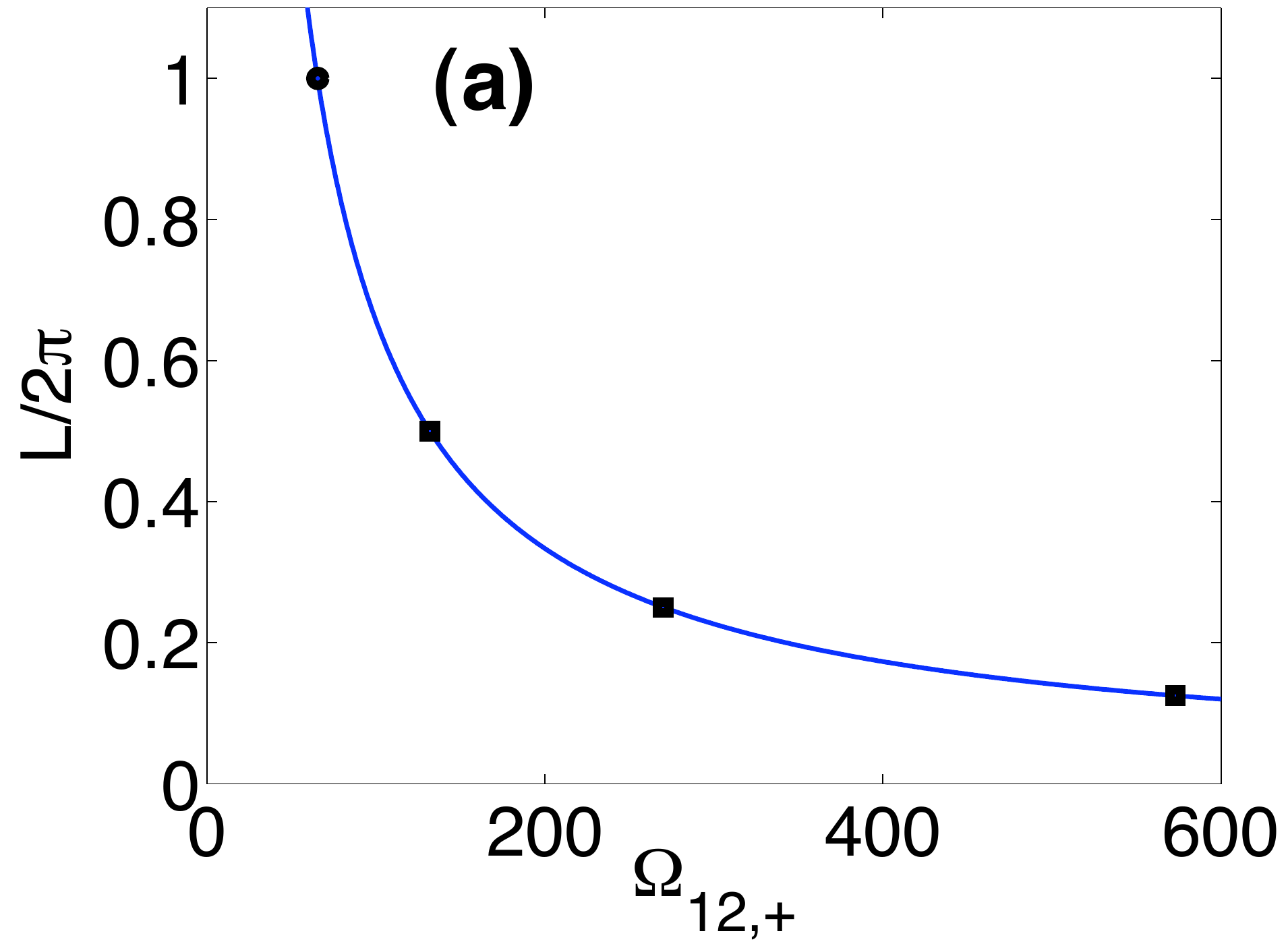}~~
\includegraphics[height=3.1cm]{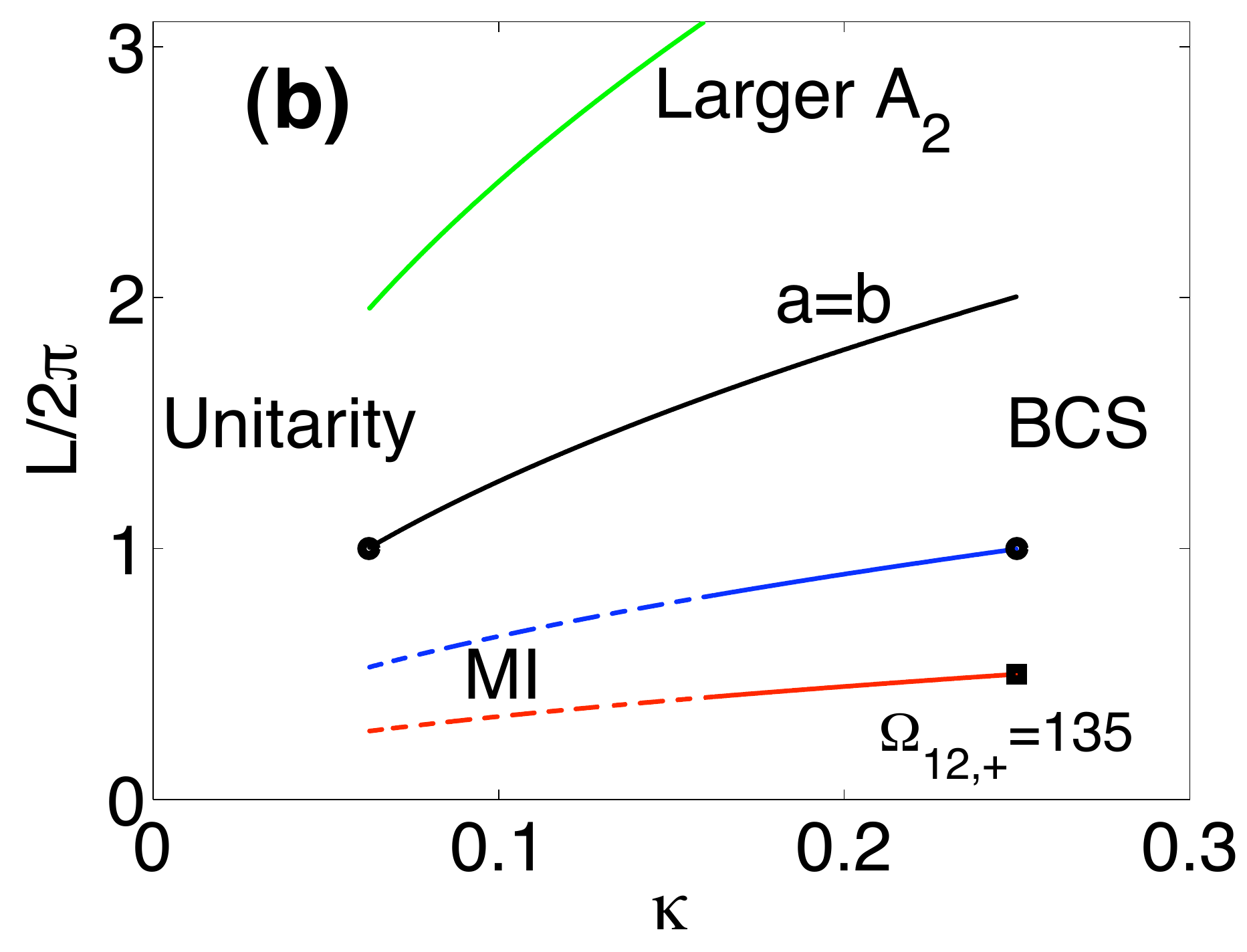}
\end{center}
\caption{(Color online) Spatial wavelengths of the Faraday patterns. (a) We use the same parameters as for Fig.~\ref{Fig1} with the left dot for $k=1$, corresponding to $\Omega_{12,+}=65.5$, and squares for $k=2,3,4$. 
(b) The case corresponding to Fig.~\ref{Fig1} (right dot) is the second (blue) curve from the bottom. Here, as well as for the lowest (red) curve [left square in (a)] where $\Omega_{12,+}=135$ ($k=2$), the condition~(\ref{StabilityCondition}) is not fulfilled in the entire domain, hence MI is possible in the left part (dashed lines). The condition~(\ref{StabilityCondition}) for not having MI can be satisfied in the entire $\kappa$ range, e.g., by choosing $A_2=2 \sqrt{20}$ larger [top (green) curve]. Finally, the special case of $a=b$ (i.e., with $ g_b^{(0)} \propto \kappa$) is illustrated with the third (black) curve, where the left dot corresponds to the case illustrated in  Fig.~\ref{Fig4}.}
\label{Fig3}
\end{figure}

\begin{figure}[tbp]
\begin{center}
\includegraphics[height=3.15cm]{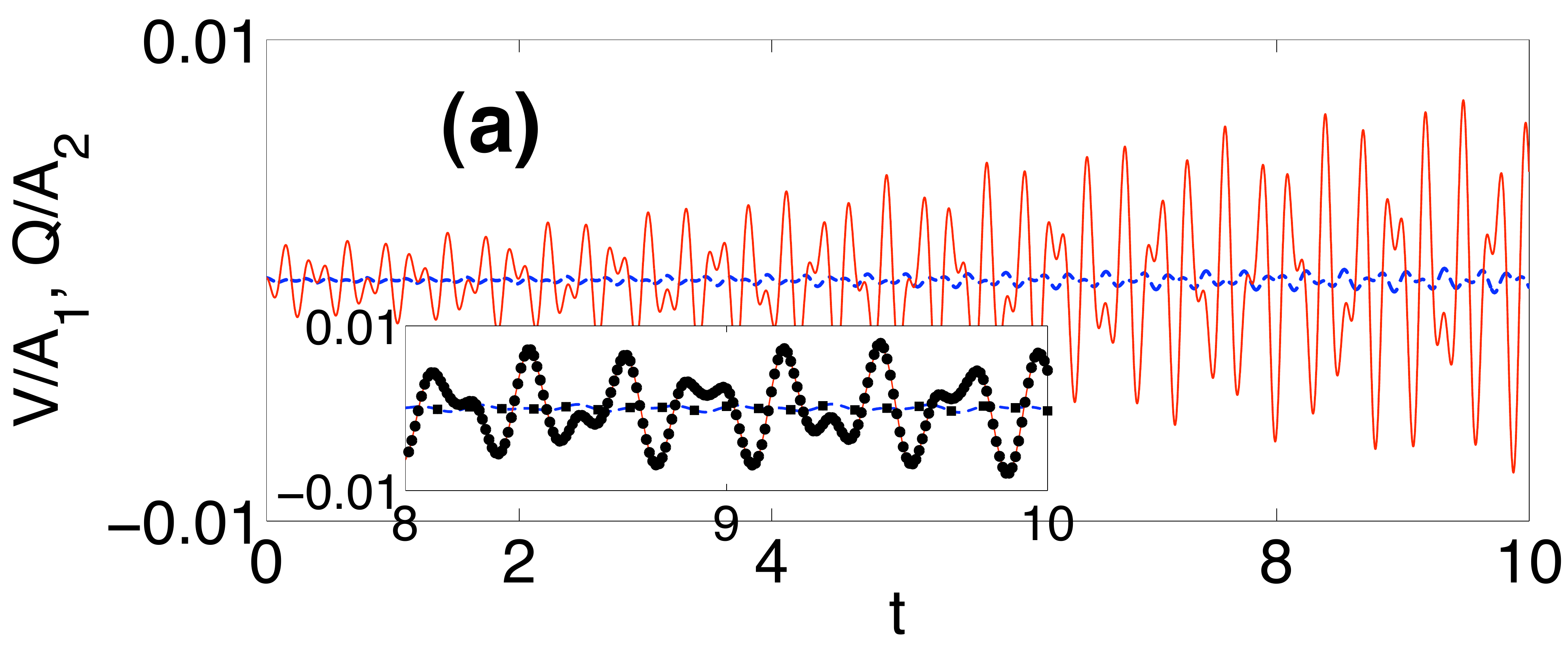}
\includegraphics[height=3.1cm]{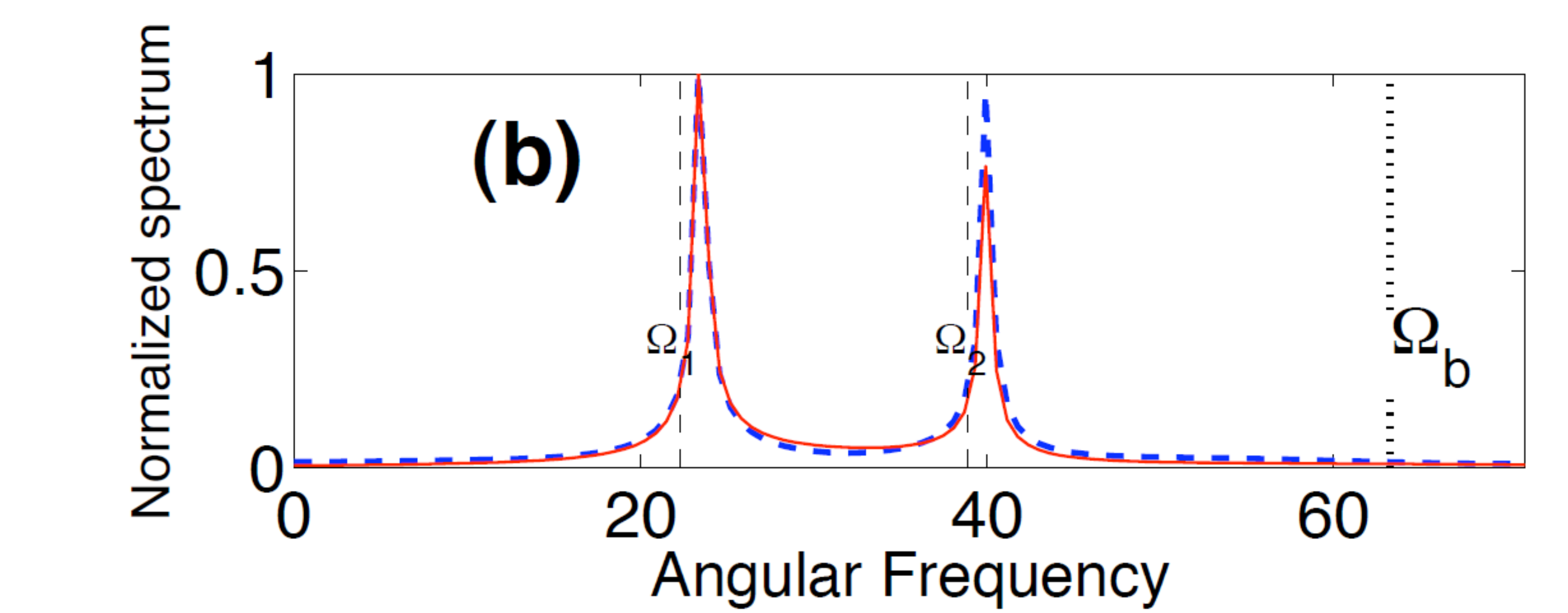}
\end{center}
\caption{(Color online) Solutions to the ODE model of Eq.~(\ref{EqForIntraMod}). (a) Amplitudes growing exponentially in time. 
The inset shows a comparison with the corresponding PDE data ($x=0$ slices of Fig.~\ref{Fig5} for selected times): black squares are for bosons and dots for fermions.
(b)~Spectrum for the modulation frequency $\Omega_b= 63.3 $ ($k=1$) and with $\Omega_{1}$ and  $\Omega_{2}$  agreeing well with Eq,~(\ref{Omega_1and2_limit_2}) in this regime. The initial conditions are the same as in Fig.~\ref{Fig1}; the parameters here are $\alpha_b=0.25, \ A_1=\sqrt{5000}, \ A_2=\sqrt{20.2}, \ g_b^{(0)} = 0.1, \ g_{12}^{(0)} = 0.8$, and $ \kappa =1/16$.}
\label{Fig4}
\end{figure}

\begin{figure}[tbp]
\begin{center} 
\includegraphics[height=3.05cm]{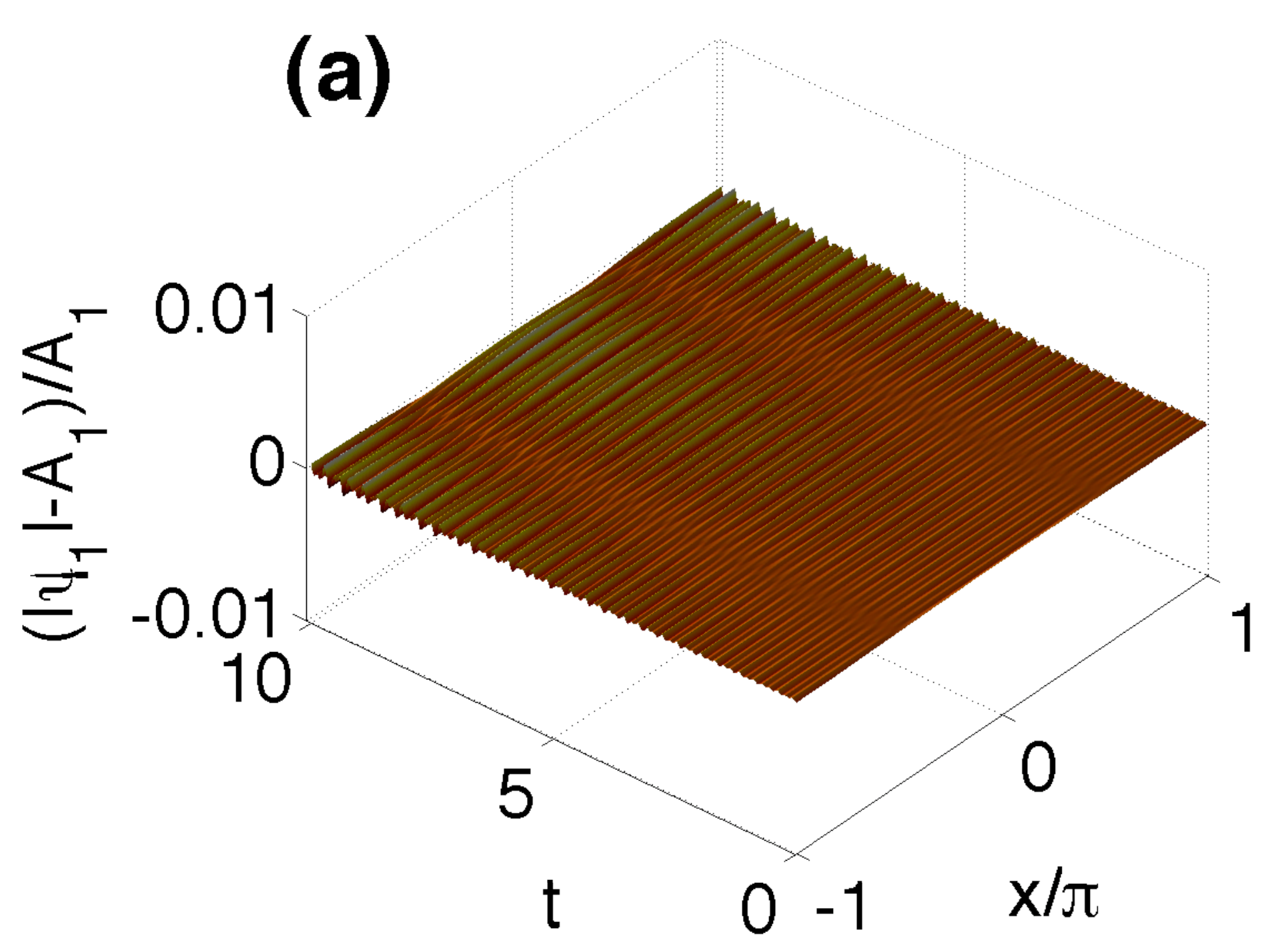}~~
\includegraphics[height=3.05cm]{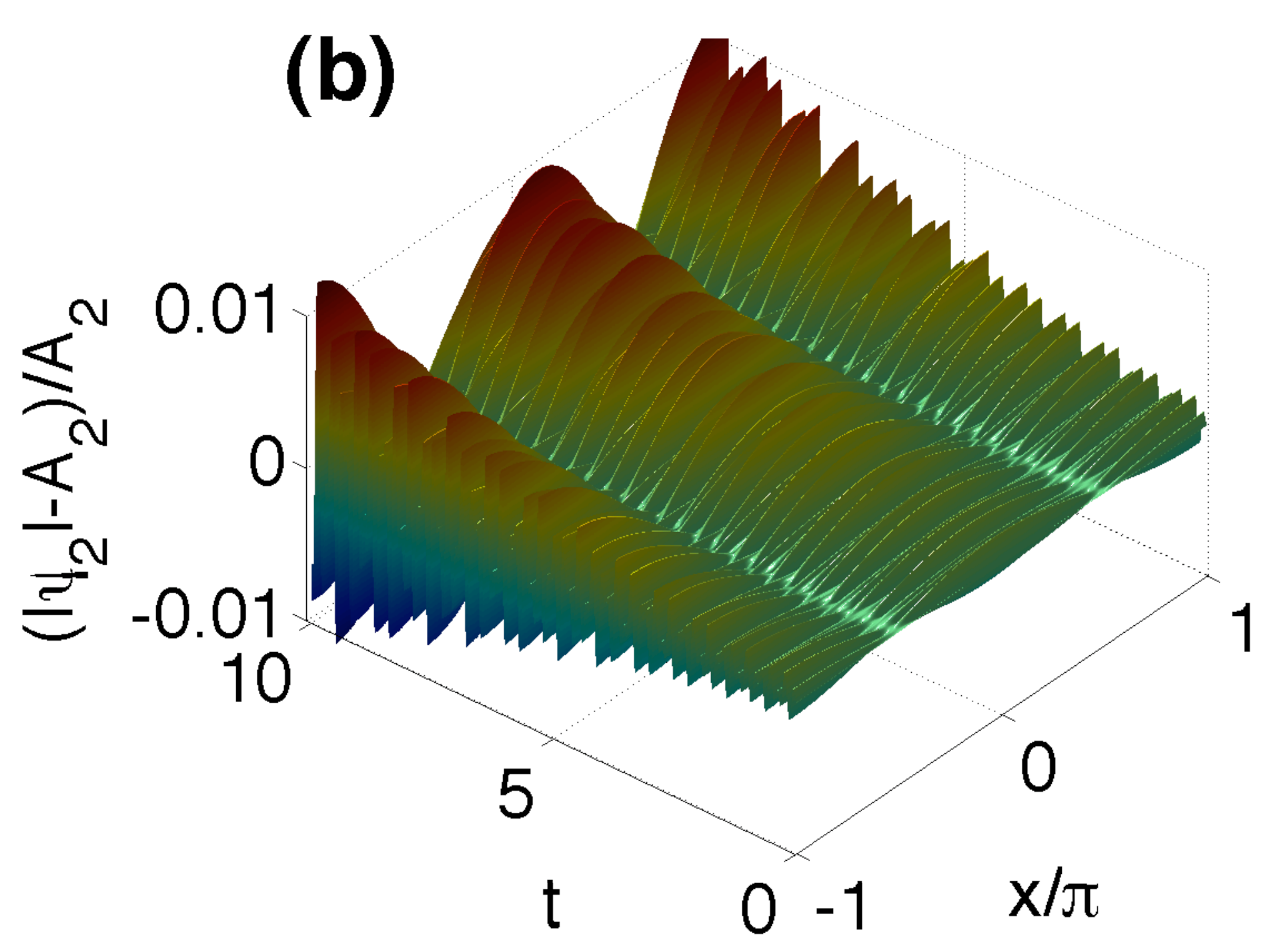}
\end{center}
\caption{(Color online) Solutions to the full PDE model of Eq.~(\ref{bfsys}) for driven nonlinearity for bosons with the modulation frequency $\Omega_b= 63.3$ ($k=1$). 
The initial and boundary conditions are the same as in Fig.~\ref{Fig2}; the parameters  are the same as in Fig.~\ref{Fig4}.}
\label{Fig5}
\end{figure}

Numerical solutions of the ODE model (\ref{EqForInterMod}) are presented in Fig.~\ref{Fig1} for parameters corresponding to the BCS regime. 
It is found that amplitudes increase exponentially in time and the behavior agrees well with the theoretical predictions. 
Using the same parameters, we perform numerical simulations of the full partial differential equation (PDE) model~(\ref{bfsys}) describing the Fermi-Bose mixture (see Fig.~\ref{Fig2}). 
Results shows quantitative agreement for the PDE and ODE models up to $t \sim 10$ [see inset of  Fig.~\ref{Fig1}(a)].
From Fig.~\ref{Fig3}, where data for the spatial wavelength $L$ of the FW versus the modulation frequency are plotted [Fig.~\ref{Fig3}(a)], it is seen how the wavelength decreases with increasing frequency.
While results for the wavenumber $k=1$ have been presented in  Figs.~\ref{Fig1} and~\ref{Fig2}, we have numerically confirmed the results for $k=2,3,4$ also with full PDE simulations [squares in  Fig.~\ref{Fig3}(a)].
In  Fig.~\ref{Fig3}(b) we show how $L$ depends on the fermionic subsystem parameter $\kappa$ within the analytic model~(\ref{L}).

The results of Eq.~(\ref{widths_a}) in practice here mean that
with the parameter values of Fig.~\ref{Fig1}, the modulation frequency can be on the order of 1\% larger (or smaller) when exciting the FW. 
This is confirmed numerically with Eqs.~(\ref{bfsys}) and~(\ref{EqForInterMod});
we have also observed a weak dependence on $\alpha_{12}$ for the maximum of the numerical $\Omega_{12}$ resonance region  (not shown).
Hence, by optimizing $\Omega_{12}$, a larger amplitude can be obtained (i.e., curve~A in Fig.~\ref{Fig6} can be moved further towards the line of the theoretical gain).

\subsection{Driven nonlinearity in the bosonic subsystem} \label{secB}

The case where only the intraspecies (boson-boson) interaction parameter is modulated is described by the system~(\ref{EqForIntraMod}).
Applying again the multiscale approach (see the Appendix), 
we have concluded  that resonances occur at $\Omega_b = 2\omega_1$ under the additional condition that $\omega_1 \simeq \omega_2$. 
Hence, from Eq.~(\ref{def_w1_w2}) with $\omega_1=\omega_2$ we now obtain
\begin{equation}
\Omega_b = 2k \sqrt{k^2 + a}, \ \  k^2 = \frac{a}{2} \left(   \sqrt{1+ \Omega_b^2/a^2} - 1 \right). \label{Omega_b}
\end{equation}
 The corresponding instability region is  bounded by the lines
\begin{equation}
\bar{\omega}_1^2 = \omega_1^2 \pm \frac{\varepsilon_0\alpha_{b}}{2}, \ \bar{\omega}_2= \bar{\omega}_1, \label{widths_b}
\end{equation}
and the maximal gain is now equal to
\begin{equation}
p_m \simeq \frac{\varepsilon_0 \alpha_{b} }{4\omega_1}.  \label{p_B}
\end{equation}

Numerical solutions of the ODE model~(\ref{EqForIntraMod}) with exponentially growing amplitudes  are presented in Fig.~\ref{Fig4} for parameters corresponding to the molecular unitarity limit. 
In the spectrum we can see peaks corresponding to the combination of two excited frequencies for both components.
These predictions are again well confirmed by PDE simulations~(Fig.~\ref{Fig5}) of the system~(\ref{bfsys}) with driven nonlinearity of bosons [see the inset of  Fig.~\ref{Fig4}(a)].
Although the bosonic component $\psi_1$ is lagging behind (i.e. lower amplitude in Figs.~\ref{Fig4} and~\ref{Fig5}) it is growing exponentially with the same rate as $\psi_2$.

\begin{figure}[tbp]
\begin{center} 
\includegraphics[height=4.2cm]{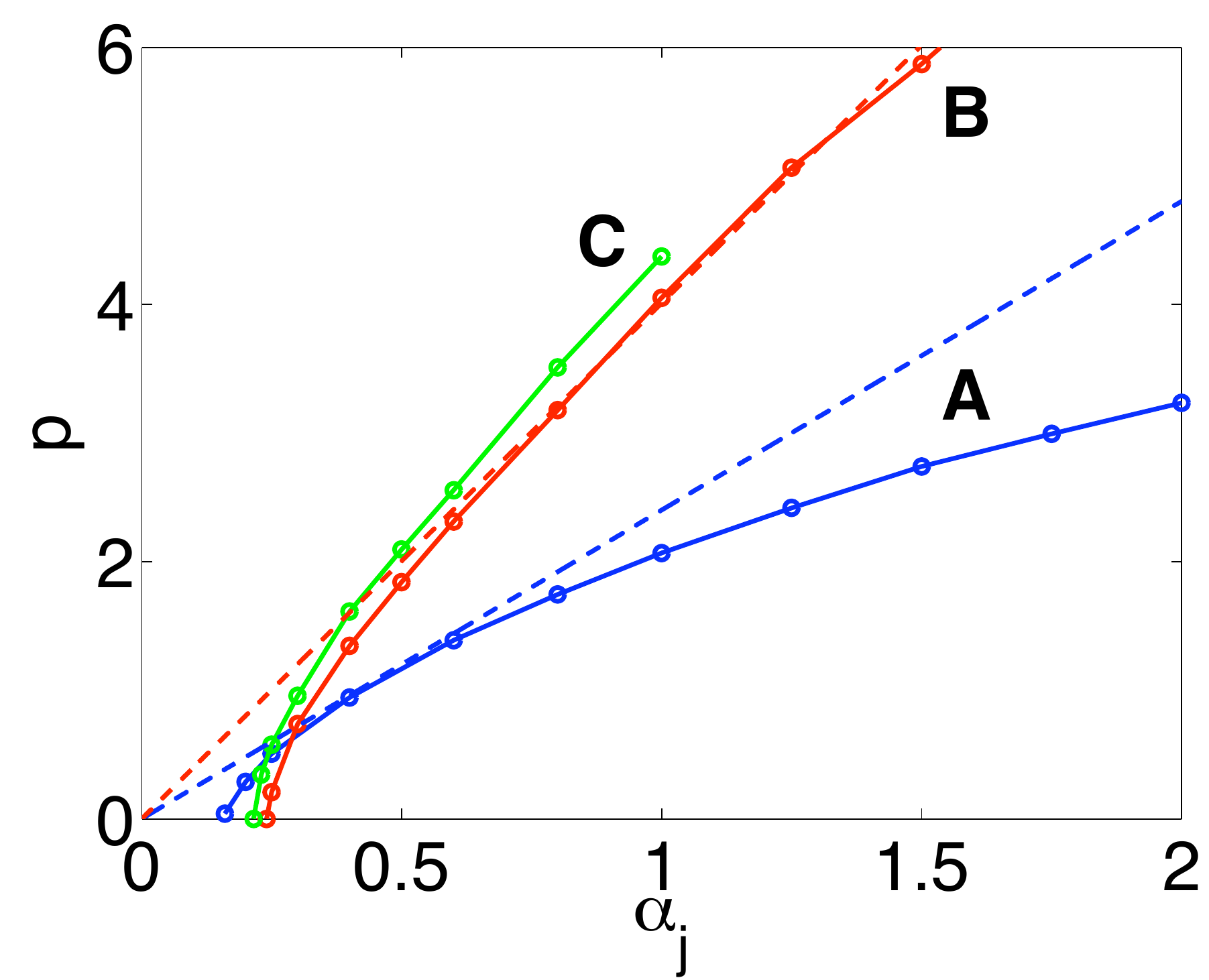}
\end{center}
\caption{(Color online) Exponential growth rates $p$ of the slowly varying amplitudes [envelope of $V \sim Q \sim \exp(pt)$ of Eq.~(\ref{TDODE})]. Solid curves connect numerical data points (circles) for the three cases: A~(Fermi-Bose modulation $\alpha_j=\alpha_{12}$), B~(Bose-Bose modulation $\alpha_j = \alpha_{b}$) and C~(super-resonance $\alpha_j =\alpha_{b} = - \alpha_{12}$).
Dashed lines shows theoretical predictions according to Eq.~(\ref{p_A}) for case A and Eq.~(\ref{p_B}) for case B.
In case A, $\Omega_{12}=65.5$, while in cases~B (C) we use $\Omega_{b}=63.3$~($=\Omega_{12}$), respectively.}
\label{Fig6}
\end{figure}

According to Eq.~(\ref{widths_b}), the modulation frequency can be on the order of 3\% larger or smaller (for the parameters of Fig.~\ref{Fig4}) when exciting the FW. 
This has been confirmed numerically with Eqs.~(\ref{bfsys}) and~(\ref{EqForIntraMod}),
although the maximum of the numerical resonance region (not shown) was found for slightly lower values of $\Omega_b$ than the theoretical estimate.
Hence, this means that curve B in Fig.~\ref{Fig6} can be made steeper by using slightly lower values for $\Omega_b$.

Finally, we can analyze also this case in terms of the symmetric (or antisymmetric) combinations $\xi_{\pm}$.
From Eq.~(\ref{EqForIntraMod}) we then have the driven coupled Mathieu-like system
\begin{equation}
\xi_{\pm ,tt} + \Omega_{1,2}^2  \xi_{\pm} +\alpha_{b} \omega_1^2 \cos( \Omega_{b} t) [  \xi_{+} + \xi_{-}]/2 =0,
\end{equation}
and in agreement with our findings above, the literature here states that $\Omega_b \simeq 2\Omega_1, \  2\Omega_2$ and $ \Omega_1 +  \Omega_2$~\cite{Indus, Hansen}.

\subsection{Driven super-resonance}

In the case of super-resonance we refer to the situation where both Bose-Bose time-modulated interactions [$\alpha_b \neq 0$ in Eq.~(\ref{TDgbANDg12})] and Fermi-Bose time-modulated interactions ($\alpha_{12} \neq 0$) are simultaneously present in Eqs.~(\ref{bfsys}) and~(\ref{TDODE}).
Comparing Eq.~(\ref{Omega_k_2}) with Eq.~(\ref{Omega_b}), we note that when $a=b$, we have $\Omega_{12,+} = \Omega_b$.
Now we change a sign in Eq.~(\ref{TDgbANDg12}) such that for $\alpha_{12} = - \alpha_b$,  for example, the two types of modulations initiate opposite phases for the two components and an increased growth rate of the Faraday waves is observed, as compared to the two distinct cases discussed before. 
To demonstrate the super-resonance numerically we used the same parameters as for Fig.~\ref{Fig4} with $\alpha_{12}=-0.25$. 
This result can also be understood from a consideration of the system (\ref{TDODE}) for symmetric and antisymmetric combinations $\xi_{+}$ and $ \xi_{-}$. 
For example, in the case when   $\alpha_{12} = - \alpha_b/2$ and $\Omega_{12}=\Omega_b$, 
we simply have the driven coupled Mathieu-like system
\begin{equation}
\xi_{\pm ,tt} +[\Omega_{1,2}^2 + \Omega_{2,1}^2  \frac{\alpha_{b}}{2} \cos( \Omega_{b} t) ] \xi_{\pm}  = -\omega_1^2 \frac{\alpha_{b}}{2} \cos( \Omega_{b} t)  \xi_{\mp}.
\end{equation}

\section{Growth of the Faraday Waves}

In general we note,  that in the regime where the linear stability analysis based on Eqs. (\ref{ansatzLSA}) and (\ref{LinearSystem}) is valid we have quantitative agreement between the full model~(\ref{bfsys}) (see Figs.~\ref{Fig2} and~\ref{Fig5}) and the ODE models~(\ref{EqForInterMod}) and~(\ref{EqForIntraMod}) (Figs.~\ref{Fig1} and~\ref{Fig4}, respectively).
We found an exponential growth in time of the oscillating amplitudes in this regime. 
For larger times the nonlinearities of the full model (\ref{bfsys}) cause a saturation of the amplitudes, while the results of the linear stability analysis become unphysical. 

In Fig.~\ref{Fig6} we show results of the exponential growth rate for Secs. IV~A-IV~C together with the theoretical results for the maximal gain derived in the Appendix.
We show results only for the wave number $k=1$; however, we note that the theoretical estimates $p_m(k)$ increase with $k$ and asymptotically approach the constant $\lim_{k \rightarrow \infty} p_m(k) = \alpha_j g_{12}^{(0)} A_1 A_2 /2$.

\section{Realistic parameters}

The Faraday waves can be observed, for example, in mixtures such as $^{41}$K-$^{40}$K and $^{87}$Rb-$^{40}$K~\cite{Wu}. 
Here we estimate effective values for the first system above since the two atomic masses are almost equal then. 
The fermion-boson scattering length $a_{12}$ can be tuned by using the Feshbach resonance techniques according to~\cite{Collective}
$$
a_{12}=a_{bg} \left( 1 + \frac{\Delta B}{B(t)-B_0} \right), 
$$
where $\Delta B \simeq 53$ G, $B_0 \simeq 541.5$ G, and $a_{bg}  \simeq 65a_0$ ($a_{bg}$ is the background atomic scattering length and $a_0$ the Bohr radius). 
By variations in time of the external magnetic field
$B(t)$ around $B_0$ we can tune the scattering length $a_{12}$.
For example, we take $a_{12}=250a_0$ and $a_b=85a_0$ and assume a length scale in the trap of $l=2.3 \ \mu$m.  
If we then consider the BCS regime ($\kappa=1/4$) and take the numbers of bosons and fermions as $N_b = 2\times 10^5$ and $N_f =2\times 10^3$, with the transverse trap frequency  being $\omega_{\perp}=1.9$ kHz, we find that for modulations with the frequency $\Omega_{12} \simeq 32\omega_{\perp}$ the spatial wavelength of the Faraday pattern is $L\approx 31 \ \mu$m.   
Correspondingly in the molecular unitarity limit ($\kappa=1/16$) the spatial wavelength is $L \approx 16\ \mu$m.

\section{Conclusion}

We have illustrated the possibility of Faraday patterns for Fermi-Bose mixtures, i.e., with atomic bosons coupled to fermions, in both the fermionic BCS regime and the molecular unitarity limit.
In particular 
we have investigated  quasi-one-dimensional superfluid FB mixtures with periodic variations in time of the Fermi-Bose or Bose-Bose interactions.
We find Faraday patterns for both cases and study their properties depending on the parameters for modulations and the system settings. 
Combining the two types of modulations can result in even larger amplitudes.
We also conjecture that Faraday waves can be observed in an atomic BEC coupled to a Tonks-Girardeau gas.

A natural continuation of this work is to  investigate Faraday patterns in FB mixtures for the two- and three-dimensional cases. 
This problem deserves a separate investigation though, since the corresponding coupled nonlinear Schr\"{o}dinger-like equations are different.

\section*{ACKNOWLEDGEMENTS}
F.Kh.A.  acknowledges partial support from the Funda\c{c}\~{a}o de Amparo \`{a} Pesquisa do Estado de S\~{a}o Paulo, Brazil, and through the Otto M\o nsted Guest Professorship at DTU.

\section*{APPENDIX}

To understand the resonances in the coupled system~(\ref{EqForInterMod}), we can use the multiscale analysis~\cite{Makhmoud}. 
Following this approach, we look for solutions where 
\begin{equation}
\omega_1^2 = \omega_{01}^2 + \varepsilon_0 a_1 +\varepsilon_0^2 a_2 + \cdots, \ \omega_2^2 = \omega_{02}^2 + \varepsilon_0 b_1 +\varepsilon_0^2 b_2 + \cdots \label{App_omega_ansatz}
\end{equation}
and correspondingly for the functions $V$ and $Q$
\begin{eqnarray}
V &=& V_0(t,T) + \varepsilon_0 V_1(t,T) + \varepsilon_0^2 V_2(t,T)+ \cdots,\nonumber \\
 Q &=& Q_0(t,T) + \varepsilon_0 Q_1(t,T) + \varepsilon_0^2 Q_2(t,T)+ \cdots, \label{App_V_Q_ansatz}
\end{eqnarray}
where $T=\varepsilon_0 t$ is a slow time. Taking the terms of each order in $\varepsilon_0$ we obtain from Eq.~(\ref{EqForInterMod}) the following system of equations up to the linear order in $\varepsilon_0$
\begin{eqnarray}
V_{0,tt} +\omega_{01}^2 V_0 &=& 0,  \ \ Q_{0,tt} + \omega_{02}^2 Q_0 =0,  \label{Appendix_Eqs_3} \\
V_{1,tt}+ \omega_{01}^2 V_1 &=& -2V_{0,tT}-a_1 V_0  + \left[ 1 +  \alpha_{12}\cos(\Omega t) \right] Q_0, \nonumber \\
Q_{1,tt} + \omega_{02}^2 Q_1 &=& -2Q_{0,tT} -b_1 Q_0 + \left[ 1 + \alpha_{12} \cos(\Omega t)  \right] V_0.  \nonumber
\end{eqnarray}
The solutions of the first two uncoupled equations of Eqs.~(\ref{Appendix_Eqs_3}) can be written in the form
\begin{eqnarray}
V_0(t,T) &=& A_0(T)\cos(\omega_{01}t) + B_0(T)\sin(\omega_{01}t),\nonumber\\
Q_0(t,T) &=& C_0(T)\cos(\omega_{02}t) + D_0(T)\sin(\omega_{02}t). \label{App_envelope_ansatz}
\end{eqnarray}

We require the absence of the resonant terms on the right-hand sides of the third and fourth of Eqs.~(\ref{Appendix_Eqs_3}). With the ansatz $\Omega = \omega_2 \pm \omega_1$ and after averaging in the fast time $t$, we obtain a system of equations for the envelope functions $A_0, \ B_0, \ C_0$, and $D_0$ of Eqs.~(\ref{App_envelope_ansatz}),
\begin{eqnarray} \label{App_lin_sys_A}
2 \omega_{01} A_{0,T} - a_1 B_0 \mp \frac{\alpha_{12}}{2}D_0=0,  \nonumber \\
-2 \omega_{01}  B_{0,T} - a_1 A_0 + \frac{\alpha_{12}}{2}C_0 =0, \nonumber \\
2 \omega_{02} C_{0,T} - b_1 D_0 \mp \frac{\alpha_{12}}{2}B_0 = 0, \nonumber \\
-2 \omega_{02} D_{0,T} - b_1 C_0 + \frac{\alpha_{12}}{2}A_0 = 0.
\end{eqnarray}
Looking for solutions of the form $A_0,B_0,C_0,D_0 \sim \exp(p t)$, i.e., for example, with $A_{0,T} \sim p A_0 /\varepsilon_0$, we find from Eqs.~(\ref{App_lin_sys_A}) the characteristic equation $p^4 + M p^2 + N = 0 $ with coefficients
\begin{equation}
M= \frac{ 2b_1^2\omega_{01}^2  +2a_1^2\omega_{02}^2 \mp \alpha_{12}^2\omega_{01}\omega_{02}  }{8 \omega_{01}^2\omega_{02}^2}, \: N = \frac{(4a_1b_1 - \alpha_{12}^2)^2}{256\omega_{01}^2\omega_{02}^2}. \label{rate}
\end{equation}
Remember that the sign $\mp$ in $M$ of Eqs.~(\ref{rate}) is for $\Omega_{12, \pm} = \omega_2 \pm \omega_1$ respectively, 
and note also that $N \geq 0$.
Hence, from $p^2 = -M/2 \pm \sqrt{M^2/4-N}$ it is seen that only $\Omega_{12, +}$ can correspond to a positive real $p$, i.e., a FW with an  exponentially growing amplitude,
while excitations for $\Omega_{12,-}$  are stable. 
The maximal exponential growth rate of the FW for $\Omega_{12,+}$ is found from Eqs.~(\ref{rate}) with $M^2 \sim 4N$ to be 
\begin{equation}
p_m \sim \sqrt{- \frac{M}{2} } \simeq  \frac{ \varepsilon_0 \alpha_{12}}{4\sqrt{\omega_1\omega_2}},
\end{equation}
which is referred to as the theoretical gain in the main text.

For experiments on FW it is important to know also the width of the instability region.
The boundaries of the unstable region can be found from inspection of~Eqs.~(\ref{rate}), which shows that at the boundary we have from $N$ that $b_1 =  \alpha_{12}^2/4a_1$ and correspondingly from $M$ we then have $a_1 = \pm \frac{ \alpha_{12}}{2}\sqrt{\omega_1/\omega_2}$ such that the frequencies to linear order in $\varepsilon_0$ obtained from Eqs.~(\ref{App_omega_ansatz}) are
\begin{equation}
\omega_1^2 = \omega_{01}^2 \pm \varepsilon_0 \frac{\alpha_{12}}{2} \sqrt{  \frac{\omega_1}{\omega_2} } + \cdots, \ \omega_2^2 = \omega_{02}^2 \pm \varepsilon_0  \frac{\alpha_{12}}{2} \sqrt{  \frac{\omega_2}{\omega_1} }  + \cdots.
\end{equation}

Analogously the system~(\ref{EqForIntraMod}) for the case of driven boson-boson interactions can be investigated
and the results are reported in Eqs.~(\ref{Omega_b})-(\ref{p_B}).
Below we sketch the derivation also for this case.

For $\varepsilon_0=0$ the first of Eqs.~(\ref{EqForIntraMod}) is a Mathieu equation
\begin{equation}
V_{tt} + [\omega_1^2+  \alpha_b \omega_1^2  \cos(\Omega  t)] V =0 , \label{Mathieu_equation}
\end{equation}
with solutions $V=A \: \textnormal{Ce}(a,q,t) + B \: \textnormal{Se}(a,q,t) $, where $a=4 \omega_1^2/\Omega^2$ and $q=- 2 \alpha_b \omega_1^2/\Omega^2$ in the standard notation of the cosine and sine Mathieu functions \cite{Abramowitz_and_Stegun}.
We now look again for solutions to Eqs.~(\ref{EqForIntraMod}) of the form of Eqs.~(\ref{App_omega_ansatz}) and~(\ref{App_V_Q_ansatz}).
In the case of $ \alpha_b \ll 1$ one can use the expansions $  \textnormal{Ce}(a,q,t) \sim \cos(\omega_{01} t ) +{\cal O} (q)$ and $  \textnormal{Se}(a,q,t) \sim \sin(\omega_{01} t ) + {\cal O}(q) $ in Eq.~(\ref{Mathieu_equation}).
In particular we then have that the system in the first two of Eqs.~(\ref{Appendix_Eqs_3}), and hence Eqs.~(\ref{App_envelope_ansatz}), applies also here and we have in the linear order
\begin{eqnarray}
V_{1,tt}+ \omega_{01}^2 V_1 &=& -2V_{0,tT} - a_1 V_0  -  \alpha_b \cos(\Omega t) V_0 + \varepsilon_0 Q_0, \nonumber \\
Q_{1,tt} + \omega_{02}^2 Q_1 &=& - 2Q_{0,tT} - b_1 Q_0 + \varepsilon_0  V_0.   \label{Appendix_Eqs_X}
\end{eqnarray}
Hence $A_0$ and  $B_0$ are not coupled to $C_0$ and  $D_0$ in the lowest order in $\varepsilon_0$ such that it is enough to consider the two envelope functions $A_0$ and $B_0$ in the remainder.
We then require the absence of the resonant terms on the right-hand side of the first of Eqs.~(\ref{Appendix_Eqs_X}). With the ansatz $\Omega = 2  \omega_1$ and $\omega_2 = \omega_1$ ($b_1=a_1$), we obtain a system of equations for $A_0$ (with $A_{0,T} \sim p A_0 /\varepsilon_0$) and $B_0$ such that the characteristic equations give 
\begin{equation}
p = \pm \frac{\varepsilon_0}{2 \omega_1} \sqrt{  \frac{\alpha_b^2}{4} -a_1^2}. 
\end{equation}
Hence the theoretical gain for $a_1 \sim 0$ is
\begin{equation}
p_m =  \frac{ \varepsilon_0 \alpha_b}{4 \omega_1}. \label{Appendix_case_B_p}
\end{equation}
Since $p$ becomes imaginary when $a_1^2 > \alpha_b^2/4$, we have the boundary $a_1 = b_1 = \pm \alpha_b /2$ and hence $\omega_1^2 = \omega_2^2 \simeq \omega_{01}^2 \pm  \varepsilon_0 \frac{ \alpha_b}{2} $.

\end{document}